\documentclass[11pt,journal]{IEEEtran}
\usepackage[utf8]{inputenc}
\usepackage{hyperref}
\usepackage{url}
\usepackage{amsmath}
\usepackage{graphicx}
\usepackage{authblk}
\usepackage{subcaption}
\usepackage{cite}
\usepackage{hyperref}

\title{Classification of Fixed Point Network Dynamics From Multiple Node Timeseries Data}
\author[1]{David Blaszka}
\author[2]{Elischa Sanders}
\author[2]{Jeff Riffell}
\author[1,3,4]{Eli Shlizerman}

\affil[1]{Department of Applied Mathematics, University of Washington, Seattle, 98195, WA}
\affil[2]{Department of Biology, University of Washington, Seattle, 98195, WA}
\affil[3]{Department of Electrical Engineering, University of Washington, Seattle, 98195, WA}
\affil[4]{corresponding author: shlizee@uw.edu}

\begin{document}
\maketitle
\begin{abstract}
Fixed point networks are dynamic networks that encode stimuli via distinct output patterns. Although such networks are  omnipresent in neural systems, their structures are typically unknown or poorly characterized. It is therefore valuable to use a supervised approach for resolving how a network encodes distinct inputs of interest, and the superposition of those inputs from sampled multiple node time series. 
In this paper we show that accomplishing such a task involves 
finding a low-dimensional state space from supervised recordings. We demonstrate that standard methods for dimension reduction are unable to provide the desired functionality of optimal separation of the fixed points and transient trajectories to them. However, the combination of dimension reduction with selection and optimization can successfully provide such functionality. Specifically, we propose two methods: Exclusive Threshold Reduction (ETR) and Optimal Exclusive Threshold Reduction (OETR) for finding a basis for the classification state space. We show that the classification space constructed upon combination of dimension reduction  optimal separation can directly facilitate recognition of stimuli, and classify complex inputs (mixtures) into similarity classes.
We test our methodology and compare it to standard state-of-the-art methods on a benchmark dataset - an experimental neuronal network (the olfactory system) that we recorded from to test these methods. We show that our methods are capable of providing a basis for the classification space in such network, and to perform recognition at a significantly better rate than previously proposed approaches.
\end{abstract}

\section{Introduction}
\label{headings}

Robust  neural networked  systems  encode   their dynamics by producing attractors
in a  low-dimensional state space. The attractors represent neural network response
 to various inputs and  reflect particular  states of the system. Such networks
are   common in neuronal systems  that process sensory stimuli, command motor systems,
or store memory
\cite{amit1992modeling,churchland2012neural,wills2005attractor}. The manifestation of attractors  
ensures robustness of  network performance, such that,  for a range of network
initializations and stimuli, it  exhibits  reliable dynamics.
The simplest type of attractors are input induced fixed points,
triggered
by    input signals (e.g., step functions) into a subset of network nodes, and as a response after  transient dynamics, a subset of the network output nodes produce a steady state pattern. In neuronal networks, these patterns are  identified as neural codes~\cite{averbeck2006neural}.
The network is considered selective when it is capable to  distinguish between stimuli  by producing distinct fixed points. In particular, the network
will produce similar fixed points   
for similar stimuli, and  distinguishable fixed points for distinct stimuli,  where similarity is defined by a   metric in a
low-dimensional space.
Such functionality is the primary principle upon which fixed-point networks
incorporate recognition and quantification of mixed stimuli. In addition, these networks are typically `robust' as they are capable of perform under high dynamic input variability, i.e., low signal to noise ratio (SNR). 

\begin{figure*}[h]
    \centering
    \includegraphics[width=\textwidth]{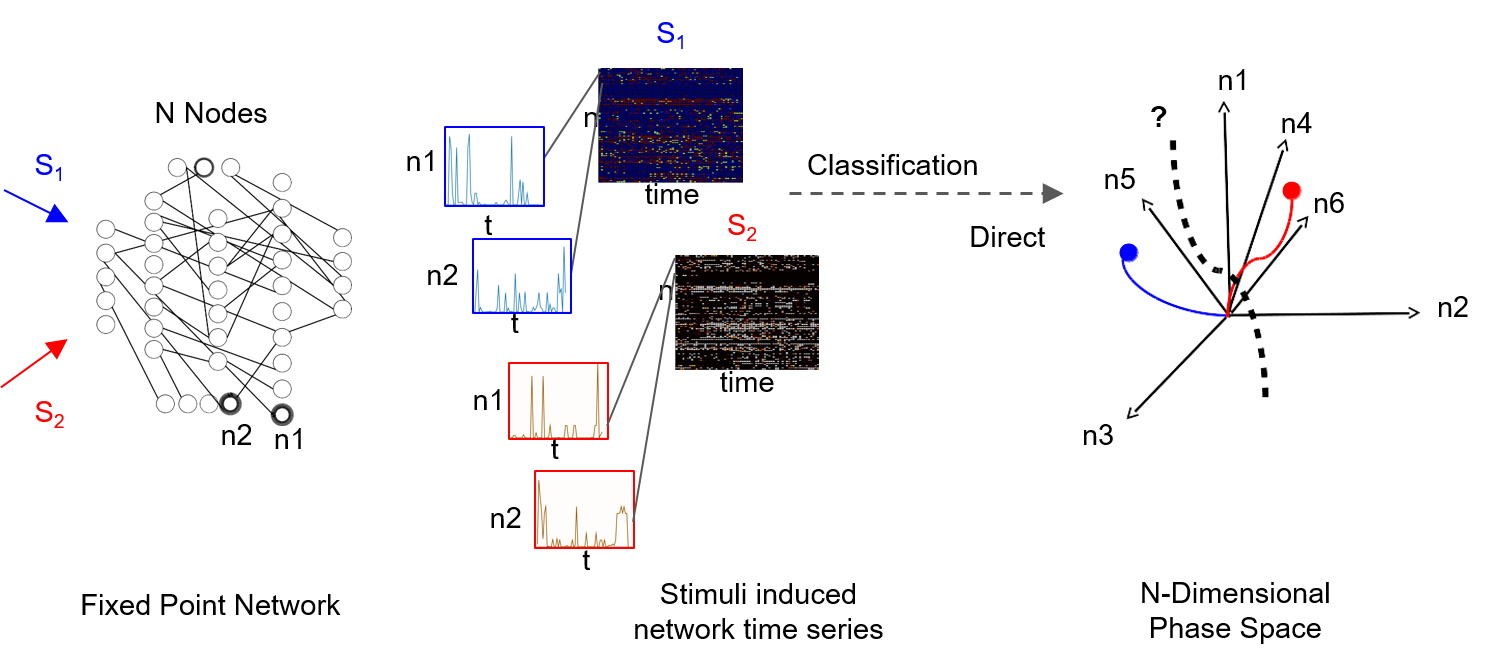}
    \caption{{\bf Supervised Classification of Fixed Point Network Responses.} Left to right: Distinct input signals, stimuli $S_i$ (here $i=2$), are injected into network's input nodes and produce fixed point dynamics in output nodes ($n1$ and $n2$ are samples of output nodes). For each stimulus output nodes timeseries are recorded in a response matrix, with dimensions of nodes $\times$ time, where each node is a row of the matrix. Response matrices to $S_1$ and $S_2$ stimuli are shown as color maps and time series of sampled nodes $n1$ and $n2$ (rows 1 and 2 of each response matrix) are shown in insets.
    As the number of output nodes increases, finding a separating hyperplane between the dynamics represented in the nodes space becomes complex as illustrated in the rightmost plot.} \label{fig:fpnetwork}
\end{figure*}

A fascinating  question is how to infer the low-dimensional state space and
 fixed points within it from sampled  time-series  data of the network.
It is a particularly relevant  problem when it is of interest  to characterize
the functionality of   black-box networks: which connectivity and  node dynamics are unknown, and probing network
response for various stimuli is the only resort. 
A structural  approach
for construction of low-dimensional state space is supervised classification.
In such an approach, a set of distinct inputs is being applied to the network
independently, and sampling of network activity, i.e., time-series of multiple
node dynamics, are being obtained. Thereby, for each input, the fragment of sampled
time-series  corresponds to a matrix with `nodes'  and `time' dimensions,
and the whole database of sampled responses corresponds to a collection of matrices (see Fig.~\ref{fig:fpnetwork}). Classification task for such a collection is to find the most informative low-dimensional state space
where the number of distinct fixed points is the number of presented distinct inputs,  and that the low-rank space could be used for examining the transient dynamics reaching these fixed points and their superpositions. Such a space would be considered optimally selective when the both the fixed points and the transient dynamics leading to the fixed points are maximally separated (orthogonal). For example, for a collection that constitutes responses to three inputs, we expect to find three basis vectors each representing a single stimulus. 

For neuronal sensory networks, the collection would typically correspond to instantaneous
firing-rates of neurons (peri-stimulus time-histograms) obtained from multi-neuron
recordings when the neural system is stimulated. A particularly intriguing system is the olfactory neuronal network in insects. Within the network, the antennal lobe, primary processing unit for olfactory neural responses, is receiving input from olfactory receptors and is able to respond with output activity which discriminates between odorants and classifies them into similarity classes. Experiments have shown that such classification is associated with behavior, and could adapt over time or after training~\cite{riffell2013neural,riffell2014flower}.  Furthermore, analysis of multi-unit electrophysiological recordings from output (projection) neurons shows that the network employs input induced fixed points to  classify the olfactory stimuli~\cite{mazor2005transient}.

Since the collection consists of matrices, a natural methodology for inference of the low-dimensional space is structuring  all sampled response matrices into a single matrix
and employing classical multivariate matrix decomposition methods, such
as Singular Value Decomposition (SVD), or sparse representations computed
by L1 minimization, that will identify dominant orthogonal patterns~\cite{shlizerman2012neural,sirovich1987turbulence}. However,
in practice, due to the ambiguity in SVD (i.e., indifference for the structure of the data and sensitivity to low SNR in the inputs), the resulting modes are mixed and non-uniquely discriminate
single inputs from the collection. Alternatively, application of dimension
reduction methods for each of the sub-matrices individually produces unambiguous
low-rank representations - however it introduces a problem of gathering the
obtained patterns into a single, joint, meaningful representation. This is
a generic problem of finding an optimal low-rank representation of multiple
instances data matrix. 

%It has applications in a wide variety of scientific
%fields: for instance, fMRI data, computer vision, and a critical problem in analysis
%of network dynamics, in particular recordings from neural systems. 

To address the problem, here we propose a  classification method that leverages
the benefits of the two approaches: orthogonality and non-ambiguity with
respect to the structure of the data. The method, called exclusive threshold reduction (ETR) operates on the low-rank representations obtained from individual applications of data reduction applied to each sub matrix, and creates an orthogonal basis. Effectively, the method exclusively associates each neuron with one of the stimuli and assigns a weight for the association. We show that such an approach can successfully separate response trajectories to the various stimuli when the response matrix is projected onto the basis. In addition, we formulate an optimization routine, called optimal exclusive threshold reduction (OETR) which allows us to achieve maximal separation of fixed points. 
For evaluation and demonstration of our methodology we have obtained a benchmark database of multi-unit time series recordings from the antennal lobe projection neurons in the \textit{Manduca sexta} moth. The recordings were obtained from the neuronal network responses to stimulation of constituents of the \textit{Datura} flower scent (single molecules and mixtures), a major food resource for the moth. The dataset is divided into a training set for construction of classification space, and a testing set to evaluate our methodology and compare with other approaches. In particular, an important test is determining how well the approach discriminates between distinct stimuli and captures similarity of various mixtures - for example, the method should be able to classify the network response to a stimulus as whether it is behavioral or non-behavioral neural code. We show that the OETR method allows such performance for various number of dimensions of classification space and mixture inputs.

\section{Results}

\subsection{Classification from Multiple Node Collection}

For supervised classification we consider a set of input nodes $\{I\}$ which receive a set of stimuli $\{S\}$. To monitor network response, we consider a set of output nodes denoted as $\{F\}$. In the training stage of supervised classification, $m$ independent stimuli $\{S_i\}_{i=1,...,m}$ are applied to the network and the timeseries of output nodes that produce dynamics attracted to stable fixed points are being recorded. Per each stimulus $S_i$, output nodes timeseries that have been recorded correspond to a matrix $[F_{S_i}]$ of dimensions $N \times T$ with rows being the nodes and columns are recording time stamps. Therefore the training stage of a supervised classification produces the collection
$C = \{[F_{S_1}],...,[F_{S_i}] ,...,[F_{S_m}] \}$, which includes multi-node times-series of network responses (see Fig. \ref{fig:fpnetwork} for an example). The goal of the classification is to find an optimal representation such that the responses to distinct stimuli in the collection are separable. Effectively, the problem is related to finding a separating hyperplane between each fixed point and its associated trajectories that approach/leave it and all other fixed points and their associated trajectories. In practice, the recordings are noisy since incorporate large $O(1)$ variability in time and space of the input.

\begin{figure}[!t]
    \centering
    \includegraphics[width=0.49\textwidth]{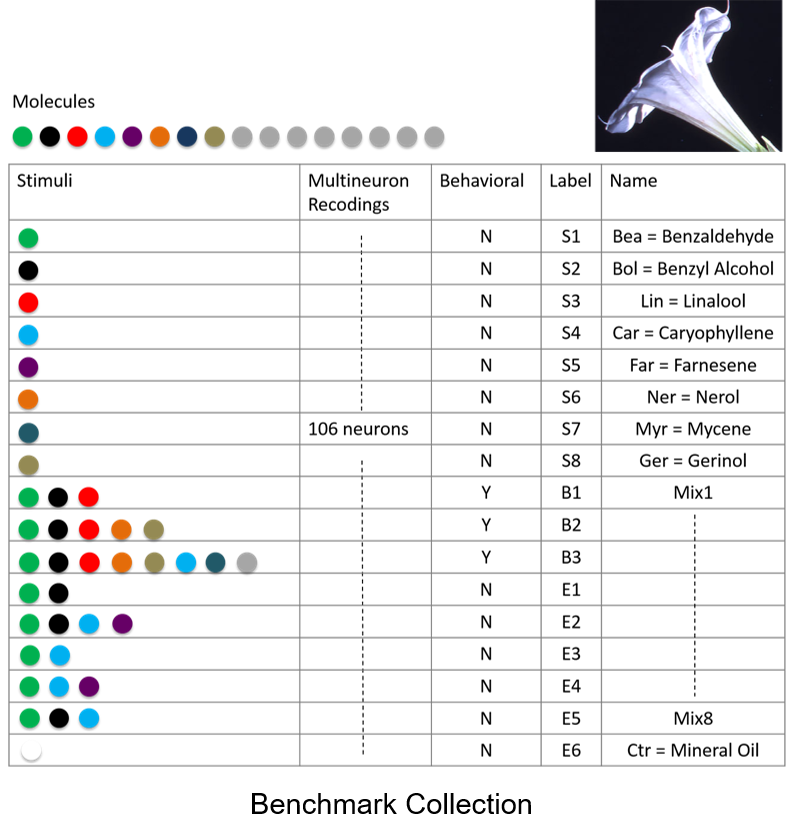}
    \caption{{\bf Benchmark data set of supervised recordings and superpositions of stimuli from the olfactory network.}  Recordings of extracellular neural responses of 106 neurons for 8 individual stimuli (odorants) that constitute the Datura scent, labeled as S1,...,S8. In addition, responses to mixed stimuli labeled as behavioral B1,..,B3, non-behavioral E1,...,E5 and control stimulus E6 were recorded. For each stimulus responses were recorded over 5 distinct trials.} \label{fig:dataset}
\end{figure}

To quantify the success of various classification methods we have established a benchmark data-set for classification and recognition. The benchmark consists of a collection of responses of neurons within a neurobiological network that exhibits fixed point dynamics. Specifically, to create the benchmark, we have obtained electrophysiological multi-neuron recordings from projection (output) neurons within the antennal lobe neuronal network, the primary olfactory processing unit in insects. It was shown that the response of this neuronal network to odor stimuli is expressed in terms of fixed point dynamics of projection neurons \cite{mazor2005transient}. 
To obtain the collection, spiking activity of multiple projection neurons (N=106) was recorded from the antennal lobe of \textit{Manduca sexta} moth subject to distinct odor molecules (odorants) and their mixtures extracted from the Datura wrightii flower, a major food resource for the moth. Since realistic response times are of order of few hundreds of milliseconds (200-400 msec), each response was recorded for 1 sec. The recordings were repeated over 5 trials with restoration intervals in between the trials. The benchmark includes responses to 8 odorants (labeled as: bea, bol, lin, car, ner, far, myr, ger), 1 control odor (labeled as: crl), and 8 mixture odors (labeled as: e2, e3, e3b, p2, p3, p4, p5, p9), see Table in Fig.~\ref{fig:dataset} for more details. Mixtures are labeled as `behavioral' - identifying that the behavioral response to these stimuli is similar to response to \textit{Datura} floral scent -, or `non-behavioral' - to denote the mixtures and odorants which do not elicit a significant behavioral response. To work with continuous time series, neural spiking activity was transformed to peri-stimulus time histograms (instantaneous firing rates) which are nonnegative values.

% Classification intro | SVM

\begin{figure*}[!t]
    \centering
    \includegraphics[width=\textwidth]{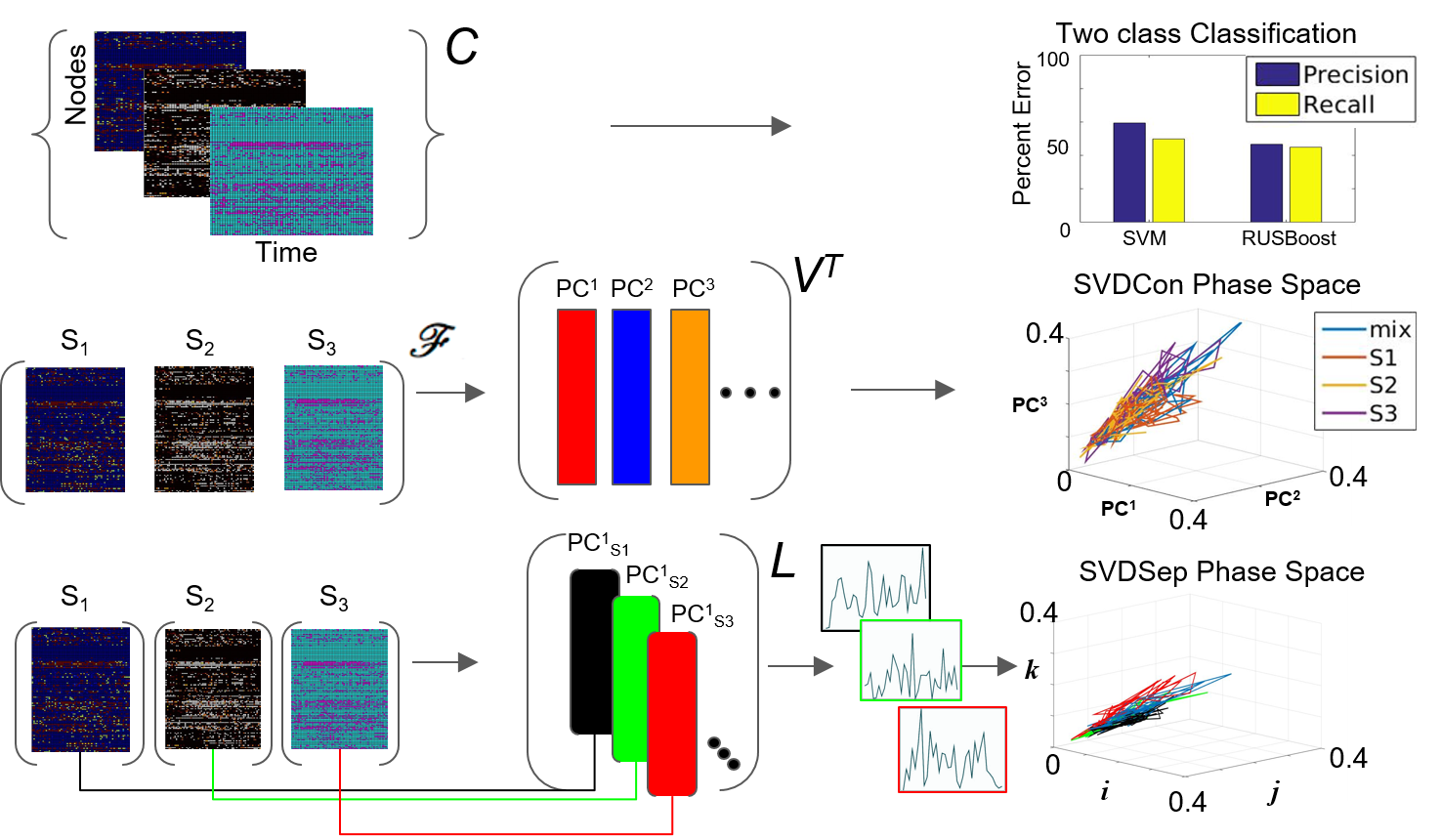}
    \caption{{\bf Possible Methods for Classification From Multiple Node Timeseries Collection} Top row: Application of SVM and RUSBoosting classifiers to perform binary classification directly on the collection $C$ in the benchmark set. Precision and recall percentages are computed and shown in the right graph. 
    Middle row: Treatment of the collection as a concatenated matrix, Eq. \ref{eq:Fconcatmatrix} -- SVDCon method. Projection of the responses onto low-dimensional space spanned by $PC^1-PC^3$ obtained from SVD on concatenated matrix shown on the right. 
    Bottom row: Treatment of the collection as a set of individual matrices, Eq. \ref{eq:Fmatrices} -- SVDSep method, and projection onto the first PC vector of each matrix SVD shown on the right.} \label{fig:collectionSet}
\end{figure*}

With the established benchmark we first demonstrate that finding a hyperplane that separates the data points is not obvious due to the high-dimensionality and variability (noise) of the data, thus precluding classification. Therefore, methods to directly identify such a separation are not expected to yield a precise classification.  To test several standard methods with the benchmark, we first used the Support Vector Machine (SVM) classifier \cite{murphy2012machine, bishop2007pattern} with a binary classification scheme for which the data was separated into two classes: responses triggered by stimuli of interest and responses triggered by other stimuli. SVM is supposed to classify the responses by finding the optimal hyperplane that separates all data points into one of the two classes (see further details on the method in the Appendix). For this binary classification problem we found the performance to be low with average classification accuracies around 50\% for various binary classes, as we show in Fig.~\ref{fig:collectionSet}. We find that both precision (percentage of true positives out of instances classified as positive) and recall (percentage of true positives out of instances expected to be positive) errors are $\approx 60\%$ and $\approx 50\%$ respectively, indicating poor performance in the usefulness of the classification (precision) and completeness of classification (recall), comparable to a random guessing strategy for the binary test sets that we have created.

A possible explanation for the failure of SVM on the benchmark is that the points are not linearly separable. We therefore tested nonlinear versions of the classifier, e.g., by using a Gaussian kernel, however they produced similar classification errors. Another hypothesis for low performance of SVM could stem from response dynamics to reside in a lower dimension than both the nodes dimension and the number of data points upon which the separating hyperplane is computed. This creates a situation where the data is being overfit. Another, obstacle could be in the form of the classes being imbalanced, i.e. when SVM classifiers are trained on an imbalanced dataset, they can produce biases towards the majority class and misrepresent the minority class. We therefore tested other techniques designed to deal with class imbalances, such as SMOTING and RUSBoost~\cite{seiffert2010rusboost} with the benchmark. However, we did not obtain significant improvement with these methods than SVM classification (Fig.~\ref{fig:collectionSet}). 
We therefore conclude that direct classification from the complete data-set is difficult and approaches that  incorporate pre-processing of the data are necessary. In particular, we identify that finding an appropriate low dimensional state space, where distinct fixed points and their associated dynamics are easily separable, could significantly simplify classification.

\subsection{Classification using Matrix Decomposition Applied to Timeseries Collection}

Since classification methods applied to the high-dimensional collection $C$ are unable to provide robust representation, an alternative approach is to represent the data in low-dimensional space. 
The structure of the collection $C$ as a set of matrices suggests that matrix decomposition could be a useful tool for finding such low-dimensional basis.
There are several possibilities to create a matrix to be used for decomposition from the collection $C$.
The first possibility is to format the collection into a concatenated matrix consisting of sub-matrices $F_{S_i}$
\begin{align}\label{eq:Fconcatmatrix}
    C=\{[F_{S_1}], [F_{S_2}], [F_{S_3}],...\} \rightarrow  \mathcal{F} = [F_{S_1}, F_{S_2}, F_{S_3},...].
\end{align}
Data reduction methods can be then applied to the matrix $\mathcal{F}$ to decompose it into vectors $\vec a_k(t_i)$ (time-dependent coefficients) and $\vec g_k(j)$ (spatial patterns/neural codes):
\begin{align}
    \mathcal{F}(t_i,j) = \sum_{k=1}^{n} \vec a_k(t_i) \vec g_k(j).
\end{align}

Our goal is to find the most informative decomposition where the
number of patterns is the number of distinct presented inputs and they span a low rank space of projections in which the dynamics of these patterns, i.e., time-dependent coefficients
$\vec a_k(t_i)$, and their superpositions could be examined. For example, for a matrix that
constitutes the responses to three inputs we expect to get three pattern
vectors each representing a single stimulus. Classical multivariate decomposition
of the matrix $\mathcal{F}$, e.g., Singular Value Decomposition (SVD) /  Principal
Component Analysis (PCA) that decompose the matrix $\mathcal{F}$ into $U \Sigma V^T$,  or sparse representations
computed by L1 minimization, can identify dominant orthogonal patterns~\cite{golub2012matrix,shlizerman2012neural}.

When computing the SVD for the matrix $\mathcal{F}$, the procedure we call SVDCon, the column vectors of orthonormal matrix $V^T$ correspond to pattern vectors. The $k$-th column vector of $V^T$, the vector $\vec g_k(j)$ (or denoted as $PC^k$ in Fig.~\ref{fig:collectionSet}), is the $k$-th pattern. Each pattern has associated time-dependent coefficients vector $\vec a_k(t_i)=\sigma_k \vec \alpha_k(t_i)$, where $\vec \alpha_k(t_i)$ is the $k$-th row vector of the orthonormal matrix $U$ and $\sigma_k$ is the $k$-th singular value, the $k$-th element of the diagonal matrix $\Sigma$. Singular values indicate the weight of each pattern vector, since they are nonnegative elements ordered as $\sigma_1 \geq \sigma_2 \geq \dots \geq 0$ and since both $\vec \alpha_k(t_i)$ and $\vec g_k(j)$ are normal vectors. To determine the relative weights of pattern vectors it is possible to define the relative energy $E_k$ of each pattern,
such that $E_k = {\sigma_k^2}/{\sum_{i=1}^n \sigma_k^2}$ which yields that the total energy is normalized $\sum_{k=i}^n E_i =1$, and each $E_k$ indicates the relative `percentage' of the energy in the pattern $PC^k$.
The distribution of $E_k$ values provides an estimate for the effectiveness of the decomposition to find a low dimensional representation. When several first $E_k$ values stand out it  identifies that the patterns corresponding to these singular values are more dominant than others and the truncation of the remainder modes would maintain a reasonable approximation of the original matrix.

We apply SVDCon to a concatenated matrix $\mathcal{F}$ generated from three response matrices,$\mathcal{F}=[F_{S_1}, F_{S_2}, F_{S_3}]$ to three distinct odorant stimuli from the benchmark set. We observe that $E_k$ values distribution is such that only the first singular value stands out ($E_1 = 0.5$) and all other are significantly smaller. Such distribution typically identifies that SVD was not able to capture the variability in data. Furthermore, examination of time dependent coefficients associated with the first three patterns shows that they are overlapping and indistinguishable (Fig. \ref{fig:collectionSet}). Effectively, these results indicate that there is discrepancy between the expected three dimensional state space and the outcome of SVDCon. 

% Joint space 

Another possibility to format the collection to a matrix form is to consider each response matrix $F_{S_i}$ separately, i.e.,
\begin{align}\label{eq:Fmatrices}
    C=\{[F_{S_1}], [F_{S_2}], [F_{S_3}],...\} \rightarrow  [F_{S_1}], [F_{S_2}], [F_{S_3}],...
\end{align}
Using this form, SVD can be applied to each matrix, procedure we call SVDSep. In this procedure $m$ sets of decompositions are obtained, where $m$ is the number of distinct stimuli. We apply SVDSep onto three stimuli collection, as for SVDCon, however here it is formatted into three separate response matrices $[F_{S_1}], [F_{S_2}], [F_{S_3}]$, and obtain 3 decompositions. We observe that each decomposition is dominated by 1 pattern ($E_1 >0.5$ for each of the three matrices).

Therefore collecting dominant pattern vectors from all decompositions into a common set could represent a projection space for the distinct responses. More generally, a procedure following SVDSep would take the first pattern vector $PC^1_{S_i}$ (i.e. most dominant mode) from each decomposition of $[F_{S_i}]$ and store them as column vectors in a library matrix $L$, which has the form: 
\begin{align*}
L = 
\begin{bmatrix}
    \vline & \vline  \\
    \vline & \vline   \\
    PC_{S_1}^1 & PC_{S_2}^
    1& \dots  \\
    \vline & \vline \\
    \vline & \vline  \\
\end{bmatrix}
\end{align*}
where there are $n$ rows accounting for every node measured in the network. Notably, however, column vectors of $L$ are taken from separate decompositions and therefore non-orthogonal to each other and do not share the same Euclidean space. Therefore projections onto the column vectors of the library $L$ are not guaranteed to be non-overlapping. Indeed, we observe that that projection of the response matrices onto $L$ for the constructed library $L$ from the benchmark experiment of three distinct odorants, does not produce separable fixed points and associated trajectories (see Fig.~\ref{fig:collectionSet}). 
The non-orthogonality of the basis warrants development of an approach to transform the library so that the projected fixed points and their trajectories are maximally separable.
\begin{figure*}[!t]
    \centering
    \includegraphics[width=0.95\textwidth]{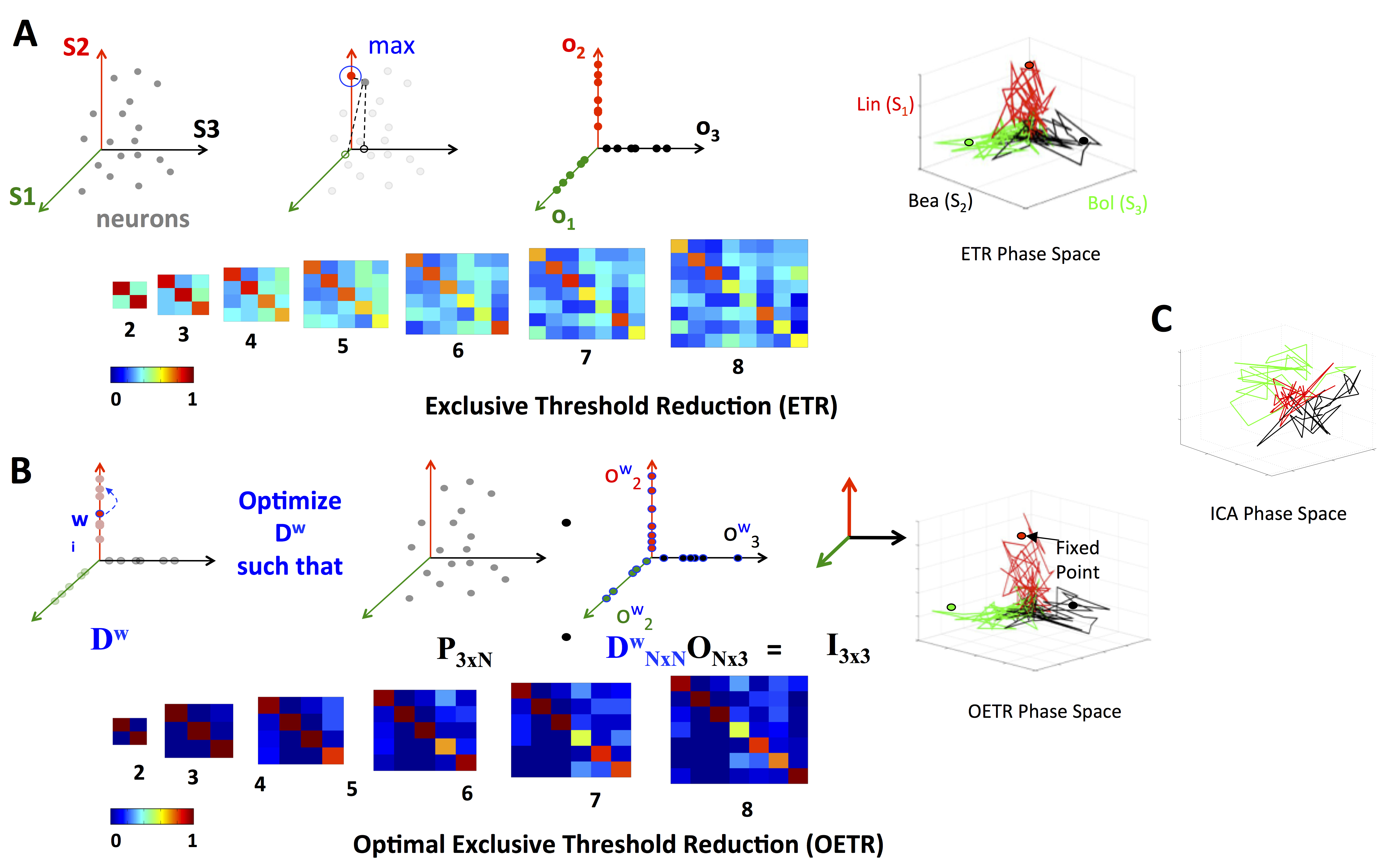}
    \caption{{\bf Exclusive Threshold Reduction (ETR) and Optimization (OETR).}
A: The exclusive threshold reduction (ETR) method applies a maximizing rule (top left) to each neuron/node in the matrix $L$ and produces the matrix $O$ where $L^T O$ defines coordinates of fixed points (bottom left). Projection of distinct stimuli trajectories onto the matrix $O$ resulting in a phase space where each stimulus trajectory is correlated with its associated axis (top right; compare with C). B: 
Optimal exclusive threshold reduction (OETR) re-weights the nodes by solving a convex optimization problem associated where $D^w$ is optimized to find the optimal weighting of $L^TD^wO = I$ (top left). The coordinates of fixed points are then defined as $L^TD^wO$ (bottom left). This further separates the transient trajectories and associated fixed points (top right). C: projection onto basis obtained by ICA method.} \label{fig:ETR}
\end{figure*} 

\subsection{Optimal Exculsive Threshold Reduction Method}
To find separation between fixed points associated with distinct stimuli, we propose a simple method that will obtain an orthogonal set of vectors from the matrix $L$. The method, called exclusive threshold reduction (ETR), operates on the nodes dimension of $L$ (rows): it selects the maximal value in each row and sets to zero all other elements in that row. The procedure is performed on each row, and once completed, leaves a new matrix with $N$ nonzero elements out of $N^2$ elements, effectively associating each node with a stimulus. For example, if for node $n_i$ the element in the third column is the maximal element then node $n_i$ is associated with $S_3$ and the weight for the association is the value of the element, see top row of Fig.~\ref{fig:ETR} for graphical illustration of the method. Therefore, the ETR method both associates nodes/neurons with stimuli, the different axes in Fig.~\ref{fig:ETR}, and assigns weights to each association. By definition, ETR guarantees an orthogonal set of vectors, i.e., an orthogonal matrix $O$ of the form 
\[ O = 
\begin{array}{c c} &
\begin{array}{c c c} PC_{S_1}^1&PC_{S_2}^1  & \dots \\
\end{array} 
\\
\begin{matrix}
n_1\\
\\
n_2\\
\\
\vdots
\\
\end{matrix}
&
\begin{bmatrix}
    &\vline & \vline &  & \vdots\\  
    \hline &\vline & \vline & &\vdots    \\
    \hline &\vline  & \vline  & & \vdots\\
    & \vline & \vline & &\vdots\\
    \dots & \dots & \dots & \dots & \dots
\end{bmatrix}=
\end{array}
\]

\[ =
\begin{cases}
    n_{ij} ,& \text{if } abs(n_{ij})= max(abs(n_i))\geq \tau,\\
    0,              & \text{otherwise},
\end{cases}
\]
where $\tau$ is the threshold value. If $\tau=0$, as in many applications, the generated matrix $O$ is a basis for the nodes space and a projection space for trajectories associated with stimuli. The matrix defines  a mapping from the high- to low-dimensional system. 
Such a mapping was used to recover network connectivity in conjunction with proper orthogonal decomposition of population model equations for the antennal lobe neuronal network~\cite{shlizerman2014data}. The mapping is supposed to group node responses and capture the exclusive features associated with each stimulus. When we test the projection of three odorant benchmark matrices onto the matrix $O$, producing time-dependent trajectories in stimulus space,  we observe that the trajectories are separate from each other. We also locate the fixed points, which coordinates are defined as the outcome of the product $L^T O$. We find their location to be close to the stimulus axis of each trajectory (Fig.~\ref{fig:ETR}A). These experiments indicate that the ETR method is effective in mapping distinct trajectories to their own axes, and can facilitate a distinct classification space.

To test the scalability of the approach we apply ETR on various subsets
of responses to odorants in the benchmark set. We start with 2 odorants and increment by one the number of included matrices to 8 odorants. Application of ETR on the latter produces 8 dimensional space (eight column matrix $O$). In these experiments we further confirm the scalable performance of the approach and its ability to produce separate fixed points and trajectories (as shown in Fig.~\ref{fig:ETR}A). 

% Convex optimization problem

While ETR turns out to be valuable in separation of trajectories, it does not ensure generically, especially when the number of matrices in the collection grows, that the fixed points are optimally separated and hence noisy trajectories are attracted to the fixed points. To optimally separate the fixed points we propose to introduce additional weights contained in the diagonal matrix $D$
\begin{align}\label{eq:Fmatrix}
D^w = 
\begin{bmatrix}
    w_{1,1} & 0 &\dots & 0   \\
    0 & w_{2,2} & \dots & 0 \\
    \vdots & \vdots & \ddots & \vdots \\
    0 & 0 &\dots & w_{n,n}   \\
\end{bmatrix}.
\end{align}
The formulation assigns weights that scale the weight of each individual node obtained by ETR to ensure that the fixed points are orthonormal (Fig.~\ref{fig:ETR}B) and thus satisfies the assumption of distinct stimuli used in supervised classification. With such scaling  fixed points coordinates are computed as $L^T D^w O$ and in order for them to be orthonormal are expected to be exclusive in the associated axis of each stimulus. We therefore require that the coordinates of the fixed points will be represented by the identity matrix, i.e., $L^TD^wO=I$. To solve for the weights, we formulate the following convex optimization problem
\begin{align}
    \min_{D^w} \Vert L^TD^w O -I \Vert_{Fr}.
\end{align} 
We denote the generalized approach of solving the optimization problem above, in conjunction with the ETR method, as Optimal Exclusive Threshold Reduction (OETR) method. It is expected to produce optimally separable projected trajectories for multiple distinct stimuli and hence the vectors of $D^w O$ are optimal axes of the classification state space. To solve the optimization problem, computational packages for convex optimization can be used. For example, we have used the {\it CVX} package implemented in MATLAB \cite{grant2008cvx}. Application of OETR to the three distinct stimuli benchmark set indeed produces more optimally separable fixed points and their associated trajectories as shown in Fig.~\ref{fig:ETR}B.

Next, we compare OETR with other approaches, such as the Independent Component Analysis (ICA), to exclude that the benchmark set is too simple problem of separation. The ICA method that we apply is information based algorithm (Infomax) particularly designed to obtain separable signals from a collection of inseparable signals, such as the library $L$, see Methods for more details on our ICA implementation and based on \cite{hyvarinen2000independent,hyvarinen1999fast,langlois2010introduction}. With ICA, our results show that projections on the obtained vectors remain to be overlapping and do not produce efficient classification (Fig.~\ref{fig:ETR}C). 

\subsection{Recognition Metrics and Classifiers based on the Classification State Space}

\begin{figure*}[!t]
    \centering
    \includegraphics[width=\textwidth]{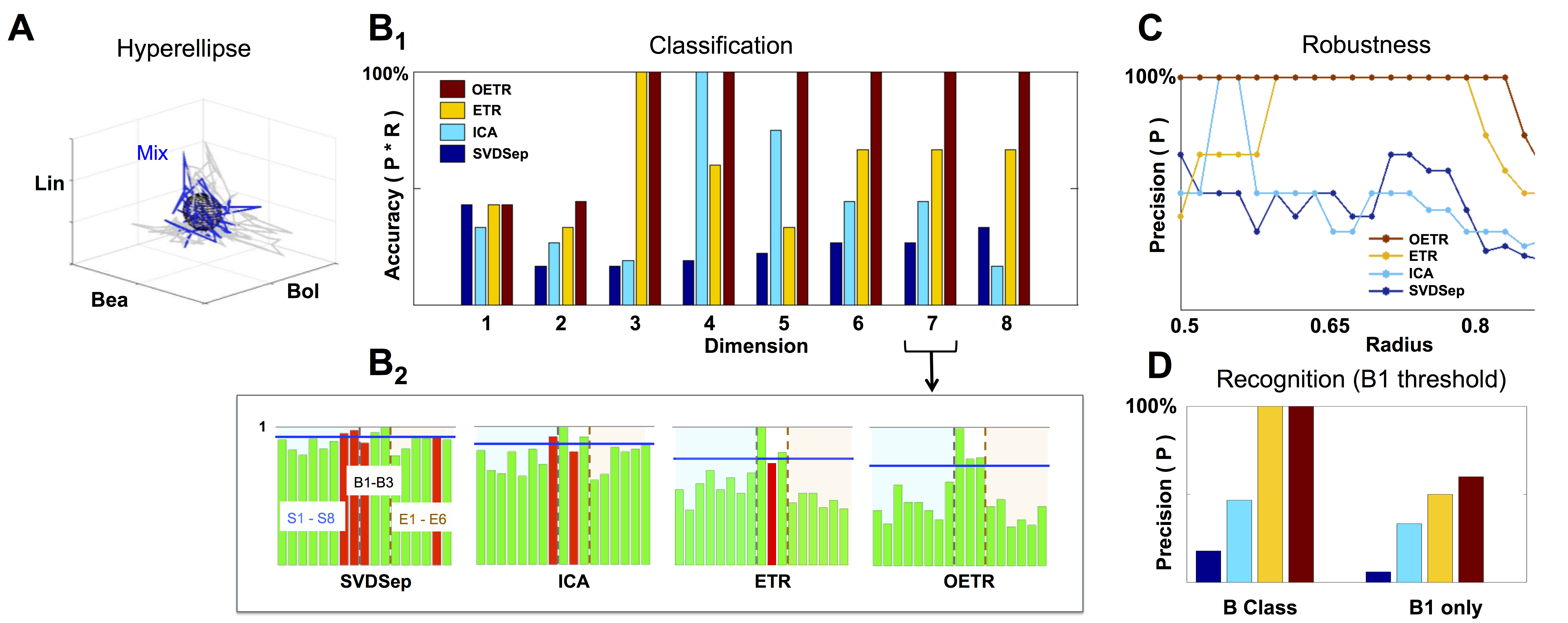}
    \caption{{\bf Classification and Recognition Employing the Classification Space.} A: ETR projection of response that corresponds to B1 (blue) and its associated 3D hyperellipse (black sphere) that is used for recognition of B1. All points that fall inside the hyperellipse are marked as ``recognition evidence" ($s_t=1$). For reference we also show the projections of responses to B1 constituents, S1, S2, S3, in gray.
    B${}_1:$ Classification accuracy into behavioral/non-behavioral classes for OETR (red), ETR (yellow), ICA (light blue), SVDSep (navy) for increasing dimension of the Classification Space. B${}_2:$ Breakdown of normalized $\left<Rec\right>$ scores per per stimulus (for 17 stimuli) for each method and $m=7$. 
    The horizontal blue line indicates the  threshold line $d$ for identification whether the response is similar to B1/behavioral. The color of the bars symbolize 
    correctness of classification: 
    correctly classified stimuli are marked by green, and incorrectly classified stimuli are marked by red. C: Precision of the each method when the radius of the classifying hyperellipse (sphere) is varied. D: Recognition precisions using B1 sphere for instantenously identifying B1 stimuli responses (right) and when when false positives from B class are allowed (left).} \label{fig:odorResults}
\end{figure*}

With the classification space constructed using ETR and OETR we consider how it can be utilized for recognition and classification of novel stimuli. These stimuli include mixtures of odorants upon which the axes of the state space were constructed and responses to odorant stimuli on trial by trial basis, where each trial is a novel stochastic realization of the trajectory than the one used for space construction. 
Since the fixed points and their associated trajectories are well separated in the classification space, we propose the convex hull metric defined by the fixed point and the associated trajectory. A simplification of the convex hull metric (and more robust) is an $m$-dimensional hyperellipse centered at the fixed point
\begin{align}
&q = (\frac{x_1 - c_1}{r_1})^2 + (\frac{x_2 - c_2}{r_2})^2 +  . . .  + (\frac{x_m
- c_m}{r_m})^2 - 1, \nonumber
\\
&s_t=
\left\{\begin{array}{lr}
        1& \text{for } q\leq 0\\
        0&  \text{for } q> 0
        \end{array}\right\}, 
\end{align}
where $m$ is the dimension of the classification state space.
The metric $s_t$ is a pointwise metric which provides for each point $x_1,...,x_m$ of the tested projected trajectory, sampled at time $t$, whether it lies inside the hyperellipse ($q\leq 0:s = 1$) or outside ($q > 0: s = 0$). Integration of the pointwise metric over a time interval of interest, typically the time of the response, e.g., for the benchmark here it is the time of the stimulus being ON ($500$ ms) provides a total score $S$ of the trajectory being located within the hyperellipse
\[
Rec = \frac{S}{T}, ~~~S = \sum_{t=t_0}^T s_t
\]
and indicates the similarity of the tested trajectory with the stimulus and its associated fixed point and hyperellipse. Normalization of the total score $S$ over the total number points provides a recognition ratio score $Rec$. Such a metric is  simple to implement and is specially designed to harness the optimal separation of fixed points in the classification space and expected to provide consistent scores for testing similarity between different trajectories in the classification space by being robust to different stochastic realizations of the trajectories (due to noise and other perturbation effects). Other metrics, measuring the time scales of convergence to the fixed point or focusing on specific properties of the trajectories are not expected to be robust for low signal-to-noise ratio (SNR) dynamics (see Fig.~\ref{fig:odorResults}).
Such properties of the dynamics are typical to multi-unit recordings from neuronal networks. 
For example, in the benchmark set SNR is less than 3 and stimuli trajectories exhibit short timescales and do not necessarily converge to the fixed points. Rather, they only approach their vicinity. 

To investigate the performance of recognition and classification using the hyperellipse metric we computed the similarity of various stimuli ($17$ stimuli from the benchmark set) with B1 mixture from the benchmark consisting of the odors: Benzaldehyde (S1); Benzyl Alcohol (S2); and Linalool (S3). This mixture was shown to include the dominant constituents of the Datura scent - a important flower nectar source of the moths - and moths stimulated with it elicit a `behavioral' response similar to stimulation by the full Datura scent, and similarly to mixtures B2 and B3~\cite{riffell2013neural}. The goal of the classification is examine the mixture trajectories and determine which are similar to B1.
Strong similarity indicates that the mixture is classified as `behavioral' as well. The goal of recognition is to decide instantaneously given a single trial whether the stimulus is B1. Notably, behavioral experiments show that similarity in responses to these different mixtures was not solely due to the concentration of the constituents making up the mixture, and therefore classification methods from neural dynamics are in need. For example, experiments show that when Linalool (component with the least concentration) is taken out of B1 mixture the new mixture becomes non-behavioral. On the other hand, adding other constituents from the Datura scent replacing Linalool does not restore the ability of the mixture to elicit behavior.

To compare between different approaches we constructed classification spaces using four methods: SVDSep, ICA, ETR and OETR for the odorants S1--S8 (see table in Fig.~\ref{fig:collectionSet}). We also varied the dimension $m$ of the classification space from $m=1 (S1)$ incrementally to $m=8 (S1,..,S8)$. For each method and dimension of the classification space we computed the score $Rec$ for each trajectory (5 trials of 17 stimuli = 85 trajectories) with respect to hyperellipse that corresponds to B1 stimulus (sphere with fixed radius). We show our results in Fig.~\ref{fig:odorResults}. In classification, for each dimension of the space, $Rec$ scores were computed for all stimuli and trials. The average $\left<Rec\right>$ score was then computed over the trials of each stimulus. The distribution of average scores $\{\left<Rec\right>\}$ for 17 stimuli is then normalized by the maximal $max\{\left<Rec\right>\}$ value. A decision line for binary classification of `behavioral' (B class) or `non-behavioral' (S or E classes) is set as the midpoint between the mean of the distribution and maximal value: $d = (\left<\{\left<Rec\right>\}\right>+1)/2$ (depicted as blue line in the inset showing distributions for $m=7$ in Fig.~\ref{fig:odorResults}). Values above $d$ are classified as behavioral while values below $d$ are classified as non-behavioral. In Fig.~\ref{fig:odorResults} the bars that correspond to erroneous classification are marked with red color and correctly classified bars are marked with green color. We show the total accuracies of the four methods (precision $\times$ recall) for each dimension as a bar plot in the middle plot of Fig.~\ref{fig:odorResults}  and also show the distributions for $m=7$ below it.

Classification results demonstrate that ETR and OETR are significant improvements over other dimensional reduction techniques, which in turn are  significant improvement over methods that do not employ dimension reduction.
All methods perform poorly when the  classification space is of $m=1$ or $m=2$ dimensions. Indeed, since B1 consists of 3 independent stimuli, the results indicate that there is not enough data for discrimination based on less dimensions than the constitutents of the tested stimulus. As the dimension increases OETR achieves perfect accuracies for $m \geq 3$. ETR achieves $100\%$ accuracy for $m=3$ and then stabilizes on lower value of $\approx 70\%$ with high precision but lower recall rates. ICA achieves $100\%$ accuracy for $m=4$ but when the dimension increases both precision and recall rates drop to values as for $m=1$ and $m=2$ dimensions. SVDSep produces constantly low accuracies for all dimensions. From the distributions, as shown for $m=7$, we get insights on the differences in the performance of the methods. In particular, two stimuli are used for validation: B1, which is expected to produce a high $Rec$ score, and E6 (control), which is expected to produce a low $Rec$ score. We observe that the scores for B1 are high across methods, however in SVDSep B1 does not cross the threshold $d$ line. For E6, only ETR and OETR successfully assign a significantly lower score to E6. As expected, individual odorants S1,..,S8 and non-behavioral mixtures E1,...E6 are far from threshold in OETR method and mostly far from threshold in ETR. The form of the distributions transforms from nearly uniform (for SVDSep) to spiked distribution with high values for B stimuli and low for other (for OETR). The latter form allows for the line $d$ to easily separate the B class from other classes. Indeed we observe that while ETR is able to perform with high precision, it is unable to recall one of the behavioral stimuli (B2). Next, we proceed and test whether the methods are sensitive to the choice of the hyperellipse. In particular, we vary the radius of B1 sphere in the range of $0.5$ -- $0.85$ for dimension $m=8$ and compute classification accuracies. We observe a clear separation between SVDSep/ICA and ETR/OETR groups of methods. We find that recall rates are stable across methods, however precision rates are sensitive. Precision rates of SVDSep and ICA are less than $50 \%$ overall and indicate that these methods are not robust. By contrast, ETR and OETR are more robust. ETR achieves $100\%$ precision for a range of $0.19$ radii and OETR achieves $100 \%$ precision in a much wider range of $0.32$.

For recognition task, we compute the $Rec$ score for $m=8$ classification space for each trial (85 trials). If the score crosses a threshold of $70 \%$ of the $Rec$ score of a target averaged trajectory (i.e. for more than 400 msec the trajectories are in the same hyperellipse) the tested trajectory is recognized as associated with the target stimulus. Using this method, we test recognition for B1 hyperellipse (sphere of radius 0.65 in our case). As in classification, our results show that OETR is more accurate than other methods. Recall and precision rates of OETR and ETR are higher than the rates of SVDSep and ICA. Interestingly, when we test for recognition of B1 stimulus only (i.e. we expect to recognize only 5 noisy trajectories of B1 stimulus as positive) precision rates do not reach $100 \%$ in any of the methods; there are other trajectories marked positively as B1 although not B1. These trajectories turn out to be mainly from B class. We show that by expanding the class of recognition and to allow for any stimulus in B class to be marked as B1. In such a case, we obtain $100 \%$ precision for OETR and ETR. These results demonstrate that noisy B trajectories appear extremely similar to each other in our recognition scheme. OETR is hence consistent with experimental observations and metrics which indicate that B class stimuli are behaviorally indistinguishable. 

\section{Conclusion}
Classification of fixed point networks responses from sampled network activity is a challenging fundamental problem, especially when the inputs into a network are highly variable and noisy. We have addressed this problem by
proposing novel supervised classification methods which find a low-dimensional representation from a collection of matrices of multiple node time-series data from the network. To test our methods we have established a benchmark database of multichannel time-series recordings from a real neural fixed point network: the antennal lobe in the {\it Manduca sexta} moth. We have shown that the OETR method that we have introduced allowed us to create a classification space that separates fixed points and their transient trajectories with high accuracy. Furthermore it allows to represent and classify noisy mixtures of stimuli. In contrast, we have shown that traditional methods such as SVM and Boosting that do not rely on dimension reduction and classical matrix decomposition and reduction methods such as SVD and ICA do not succeed at classifying fixed point networks dynamics with comparable performance to OETR. 

Recordings from the olfactory network are valuable benchmark for testing classification. The
network is known for its robustness in discrimination of scents composed of numerous distinct molecules in a highly dynamic environment. It has been established that the network employs fixed points for odor representation that are being read and processed by higher layer olfactory processing units (mushroom body in insects).
In this respect, the ETR and OETR methods that we have proposed are simple to implement since they rely on dominant mean pattern (first SVD mode) associated with each distinct odorant, maximum rule and linear weights. Such components are natural to neural networks. Thereby application of the proposed methods could help in understanding the processes that higher layer units employ to instantaneously read information from fixed point networks. Here we focused on fixed point networks. These networks are fundamental and simplest attractor networks. It is thereby  plausible that our methodology will lay the foundations for future work to extend the methodology to other more complex attractor networks such as limit cycle networks, lines of fixed points networks that are also ubiquitous networks in neural systems.

\section{Methods and Procedures}

The benchmark collection and code that implements the various methods is available online on GitHub: \hyperref[https://github.com/shlizee/fpnets-classification]{https://github.com/shlizee/fpnets-classification}.

Below we include further information on the algorithms that we have applied and compared with ETR and OETR methods.

\subsection{Class Imbalances}
If the error or misclassification percentage is the only metric taken, learning
algorithms like SVM appear to work well, but this is not an accurate result.
The reason the error is so low, is because the first class is not classified
at all and the second class is perfectly classified. Therefore, different
tactics are needed to work with imbalanced data. The first tactic, is to
change the performance metric considered. Two informative measures are {\it
precision} (the positive predictive value, or, what fraction of the labeled
class actually are in the class) and {\it recall} (the sensitivity, or, what
fraction of the relevant class was retrieved). Precision ($P$) and recall ($R$) are defined
as follows:
$P = \frac{tp}{tp+fp},~R = \frac{tp}{tp+fn}$
where {\it tp} is true positive, {\it fp} is false positive, and {\it fn}
is false negative. Another tactic, is re-sampling the data set. This includes,
oversampling the class with fewer instances to try and achieve an equal balance
between the classes, and under sampling the data to try and reduce the instances
in the larger class. However, over- or under sampling can cause biases in
the data as either the model is overfit, as in the case of over fitting, or,
in the case of under sampling, information is lost. Another technique to improve
classification performances when dealing with class imbalances, is boosting.
Boosting can improve the performance of weak classifiers, such as decision
trees, by re-sampling the training data by its assigned weights. Boosting
techniques constantly tweak weak learners until they converge towards a strong
learner. 

\subsection{SVM Binary Classification from Raw Data}

In our application of SVM approach, a hyperplane $H$ is created
to divide the classes into their proper labels (-1 or 1) and is designed to give the largest margin between these classes. The hyperplane
is then constructed as, $w^T x_i + b = 0$, where $\frac{1}{\Arrowvert w \Arrowvert}$
is the distance to $H_+$ and $H_-$ and $w$ is the vector of weights for the features in $x$. In order to obtain the optimal hyperlpane,
$\frac{1}{\Arrowvert w \Arrowvert} $ must be maximized, or $\Arrowvert w
\Arrowvert $ must be minimized. Thus, the Maximum Likelihood Estimator (MLE) for SVMs is found by minimizing
%\begin{align}
    $\frac{1}{N}\sum_{i=1}^N max\{0,1-y_i w^T x_i\} + \lambda \Arrowvert w
\Arrowvert^2$,
%\end{align}}
where $y$ is the class label and $\lambda$ is the penalty on the weights $w$. 

\subsection{RUSBoosting}
RUSBoosting is a hybrid method of random undersampling and boosting, which is designed to improve performance when classifying imbalanced data. RUSBoost randomly removes instances of the majority class to try and equalize it with the minority class. RUSBoost has been shown to work well, as well as, perform simply and speedily. Seiffert et al. \cite{seiffert2010rusboost} demonstrated that RUSBoosting shares similiar performances to SMOTEBoosting, which is a common algorithm for managing class imbalances, but does so faster and more simply. Given a training set $S$ of examples $(x_1 ,y_1)...(x_n ,y_n)$ with a minority class $y^\tau \in Y, |Y| = 2$. A new set $D'$ is generated from random undersampling, and a weak learner (i.e. a decision tree) is called to create a hypothesis $h_t$. The pseudo loss, for $S$ and $D_t$, is then calculated using $\epsilon_t = \sum_{(i,y):y_i \neq y} D_t(i)(1-h_t (x_i ,y_i ) + h_t (x_i ,y_i )) $.

The weight update parameter is then derived with: $ \alpha_t = \frac{\epsilon_t}{1-\epsilon_t}$. Normalize this and the final hypothesis $H(x) = argmax_{y \in Y} \sum_{t=1}^T h_t (x,y)log\frac{1}{\alpha_t}$.

\subsection{ICA}
We are comparing our methods with the Infomax ICA algorithm which given observed components, in our case the matrix $L^T$, finds a linear transformation $A$ from $L^T$ to $S$, i.e. $S = A L^T$. The matrix $S$ is decomposed into
independent components that minimize mutual information.  The assumption of the ICA model is that the components in $L$ are statistically independent. We are using the implementation of the infomax based on \cite{bell1995information}, in which a maximum likelihood estimation is found by using the log-likelihood in the form
$ LH = \sum_{t=1}^T \sum_{i=1}^n log f_i(w_i^T S(t)) + T log \vert det(A) \vert $
where $f_i$ is the density functions of the columns of $L^T$. This is equivalent to the Infomax Principle, $M_2 = H(\phi_1 (w_1^Tx),...,\phi_n (w_n^Tx))$ where $\phi_i$ are non-linear scalar functions. To solve the Infomax algorithm we used Python software package  \cite{
}.

\subsubsection*{Acknowledgments}
The work was supported by the National Science Foundation under Grant Nos. DMS-1361145 (ES and JAR), and IOS-1354159 (JAR), the Air Force Office of Sponsored Research under Grant No. FA95501610167 (JAR and ES), and Washington Research Foundation Innovation Fund (ES). The authors would also like to acknowledge the partial support by the Departments of Electrical Engineering, Applied Mathematics and Biology at the University of Washington.

\bibliographystyle{plain}

\end{document}